\documentclass[aps,prb,twocolumn,superscriptaddress]{revtex4-2}
\usepackage{graphicx}
\usepackage{rotating}
\usepackage{wrapfig}
\usepackage{amsmath}
\usepackage{bm}
\usepackage{upgreek}
\usepackage{comment}

\begin{document}


\title{Double-loop hysteresis of multisite dilute Sr(Y$_{1-x}$Dy$_x$)$_2$O$_4$ single crystal Kramers paramagnets: electron-phonon interaction, quantum tunneling and cross-relaxation}


\author{B.~Z.~Malkin}
\email{Boris.Malkin@kpfu.ru}
\author{R.~V.~Yusupov}
\email{Roman.Yusupov@kpfu.ru}
\thanks{Corresponding author}
\author{I.~F.~Gilmutdinov}
\author{R.~G.~Batulin}
\author{A.~G.~Kiiamov}
\author{B.~F.~Gabbasov}
\author{S.~I.~Nikitin}
\affiliation{Institute of Physics, Kazan Federal University, 420008 Kazan, Russian Federation}
\author{B. Barbara}
\affiliation{Institut N\'{e}el, CNRS/UGA UPR2940 and Universit\'{e} Grenoble-Alpes, 25 Avenue des Martyrs BP 166, 38042 Grenoble Cedex 9, France}


\date{\today}

\begin{abstract}
Experimental and theoretical studies of the dynamic magnetization in swept magnetic fields of the orthorhombic SrY$_2$O$_4$ single-crystals doped with the Dy$^{3+}$ Kramers ions (0.01 and 0.5 at.\%) with natural abundances of even and odd Dy isotopes are presented. Impurity ions substitute for Y$^{3+}$ ions at two nonequivalent crystallographic sites with the same local $C_s$ symmetry but strongly different crystal fields. Well pronounced double-loop hysteresis is observed at temperatures 2, 4, 5 and 6~K for sweeping rates of 5 and 1~mT/s. The microscopic model of spectral, magnetic and kinetic properties of Dy$^{3+}$ ions is developed based on the results of EPR, site selective optical spectra and magnetic relaxation measurements. The derived approach to the dynamic magnetization in the sweeping field based on the numerical solution of generalized master equations with time-dependent transition probabilities induced by the electron-phonon interaction, quantum tunneling and cross-relaxation allowed us to reproduce successfully the evolution of the hysteresis loop shape with temperature, sweeping rate and concentration of paramagnetic ions.
\end{abstract}


\maketitle

\section{INTRODUCTION}
Much attention has been paid to studies of the macroscopic quantum tunneling of magnetization of different Single Molecule Magnets (SMM) containing Dy$^{3+}$ ions (\cite{goodwin2017molecular,ding2018field,ortu2019studies,li2020inconspicuous,westerstrom2021precise} and references therein). A variety of hysteresis loops and their transformations, in particular, from single-loop to double-loop evolution with temperature or a sweeping rate of an external magnetic field was observed. A peculiar feature of Dy compounds is the existence of odd and even Dy isotopes with comparable natural abundances. A conventional analysis of the quantum tunneling at anticrossings of hyperfine sublevels of the ground Kramers doublet in the energy spectra of odd isotopes allows to understand, at least, qualitatively the experimental data. However, similar hysteresis loops were observed in the samples isotopically enriched with the even $^{164}$Dy isotope \cite{pointillart2015magnetic,kishi2017isotopically,ruan2023hilbert}. The step-wise loops are observed even in magnetically diluted samples though anticrossings of sublevels of the electronic Kramers doublet in the sweeping field can be induced only by the external or intrinsic (dipolar or superhyperfine) transversal magnetic field. Understanding the nature of the experimental findings in strongly diluted Kramers rare-earth electronic systems remains a challenging problem.

We report here on the experimental and theoretical studies of dynamic magnetization in dysprosium doped SrY$_2$O$_4$ single crystals. Complex oxides SrR$_2$O$_4$  (where R is Y or a rare-earth (RE) ion) possess a specific quasi-one-dimensional structure of orthorhombic symmetry with the space group $Pnam$ \cite{muller1968kenntnis,karunadasa2005honeycombs}. The unit cell contains 4 formula units, the lattice constant $c = 0.341$~nm is about three times less than the lattice constants $a = 1.007$~nm and $b = 1.191$~nm, and each sublattic is formed by ionic chains running along the $c$-axis. The impurity Dy$^{3+}$ ions substitute for Y$^{3+}$ ions at two nonequivalent crystallographic sites Y1 and Y2 with the point symmetry group $C_s$.

The physical properties of magnetically concentrated crystals SrDy$_2$O$_4$ were investigated earlier with magnetometry \cite{hayes2012magnetisation}, heat capacity measurements \cite{cheffings2013magnetic}, neutron scattering \cite{fennell2014evidence,petrenko2017evolution,gauthier2017absence,gauthier2017field}, ultrasound \cite{bidaud2016dimensionality} and muon \cite{gauthier2017evidence} spectroscopies. In the absence of an external magnetic field,  magnetic ordering in this compound was not observed down to the lowest experimentally accessible temperatures \cite{hayes2012magnetisation}. Strong magnetic anisotropy and coexistence of fast and slow fluctuations of short- and long-range magnetic correlations in external magnetic fields were discovered. The observed manifestations of a classic spin-liquid behavior of SrDy$_2$O$_4$ are undoubtedly related to the specific single ion spectral and kinetic properties of the dysprosium subsystem.

Our results of low temperature dynamic magnetization measurements in swept magnetic fields reveal that the strongly diluted paramagnet SrY$_2$O$_4$:Dy$^{3+}$ is the first example of an inorganic Kramers rare-earth compound exhibiting SMM-like properties, similar to the first non-Kramers rare-earth compound LiYF$_4$:Ho$^{3+}$ \cite{giraud2001nuclear}. However, in contrast with a single stable holmium isotope $^{165}$Ho with a nuclear spin $I = 7/2$ and the dominant role of the nuclear driven quantum tunneling of the magnetization in LiYF$_4$:Ho$^{3+}$, dysprosium has several stable even isotopes (56.2~\%) along with odd isotopes $^{161}$Dy ($I = 5/2$, 18.9~\%) and $^{163}$Dy ($I = 5/2$, 24.9~\%). 

We present here an original approach to simulation of the dynamic magnetization in a swept magnetic field for different temperatures, sweeping rates and concentrations of paramagnetic ions accounting for the electron-phonon interaction, cross-relaxation and Landau-Zener-Stückelberg (LZS) non-adiabatic quantum tunneling \cite{landau1932collisions,zener1932non,stuckelberg1932theory}. The measured low-temperature dynamic magnetization and magnetic field dependencies of relaxation rates are presented in Section~\ref{Sec2}, while Section III describes modeling of the observed hysteresis loops. The results of spectroscopic studies (electron paramagnetic resonance (EPR) and optical spectra) are presented in Supplemental Material \cite{supp} containing necessary references \cite{nikitin2023syo_Ho,fuller1976nuclear,carnall1989systematic,erdos1972electronic,freeman1962theoretical,clementi1964atomic}. The article ends with the Conclusions.

\section{\label{Sec2} EXPERIMENTAL DETAILS AND RESULTS}
In the present study, the magnetic and spectral characteristics of impurity Dy$^{3+}$ ions in Sr(Y$_{1-x}$Dy$_x$)$_2$O$_4$ ($x = 10^{-4}$ and $x = 5 \cdot 10^{-3}$) single crystals grown by the floating zone method \cite{supp,malkin2015magnetic} were measured by means of the low temperature EPR and site-selective laser spectroscopies. The analysis of the registered spectra based on the calculations of crystal-field (CF) parameters in the framework of the semi-phenomenological exchange charge model (ECM) \cite{malkin1987crystal,rousochatzakis2005master} allowed us i) to associate the observed EPR signals with exact quantum transitions between the sublevels of the CF ground state doublets  and ii) to assign spectral lines in selectively excited optical spectra to transitions between well-defined CF sublevels of the ground and several excited multiplets of the impurity Dy$^{3+}$ ions at Y1 (Dy1) and Y2 (Dy2) lattice sites. As a result, we obtained the total sets of self-consistent single-ion spectral and magnetic parameters  (two sets of 15 CF parameters (Table S3 in \cite{supp}), g-factors (Table S1 in \cite{supp}) and hyperfine coupling constants), the energy level patterns (Table S2 in \cite{supp}) and the corresponding electronic and electron-nuclear wave functions in external magnetic fields. This has served a base for a detailed modeling and interpretation of the measured dependencies of the dynamic magnetization on temperature, sweeping rate of the magnetic field and concentration of Dy$^{3+}$ ions.

It is important to underline here some specific differences between the single-ion properties of the Dy1 and Dy2 centers in SrY$_2$O$_4$:Dy crystals. Possessing the same $C_s$ point symmetry,  due to differences in the location of nearby oxygen ions (see Fig.~S1 in \cite{supp}), these centers reveal different magnetic anisotropies (of easy-axis and easy-plane type at Y1 and Y2 sites, respectively) and strongly different energy gaps (energy barriers in the relaxation processes) between the first excited and the ground CF doublets, $E_2-E_1 = 68$~K and 300~K for Dy1 and Dy2, respectively. The maximum principal value of the $g$-tensor of the ground doublet of Dy2 ions equals $g_2 = 19.28$ ($g_1$ and $g_3$ are less than 0.1) with the principal direction in the $(ab)$-plane slightly tilted ($\sim \pm 9^{\circ}$ for magnetically nonequivalent ions) from the $b$-axis contrary to Dy1 ions with the maximum $g$-factor $g_3 = 13.6$ along the $c$-axis. The discovered magnetic hysteresis is the result of long spin-phonon relaxation times of the quasi-Ising-type Dy2 ions caused by the large energy gap mentioned above.   

Magnetization measurements were carried out using the vibrating sample magnetometer (VSM) with the option of the PPMS-9 universal system (Quantum Design, USA). Magnetization ($M$) was measured as a function of a magnetic field $B_b$ applied along the $b$-axis. In order to assess the irreversible nature of magnetization, measurements were carried out in increasing and then decreasing field 
with a constant speed $v = dB/dt$. The equilibrium magnetization was measured at a reduced number of field values with setting the target field, then leaving a sample to relax (typically for 5 minutes) and then taking the magnetization measurement.

\begin{figure*}
\includegraphics[width=0.7\linewidth]{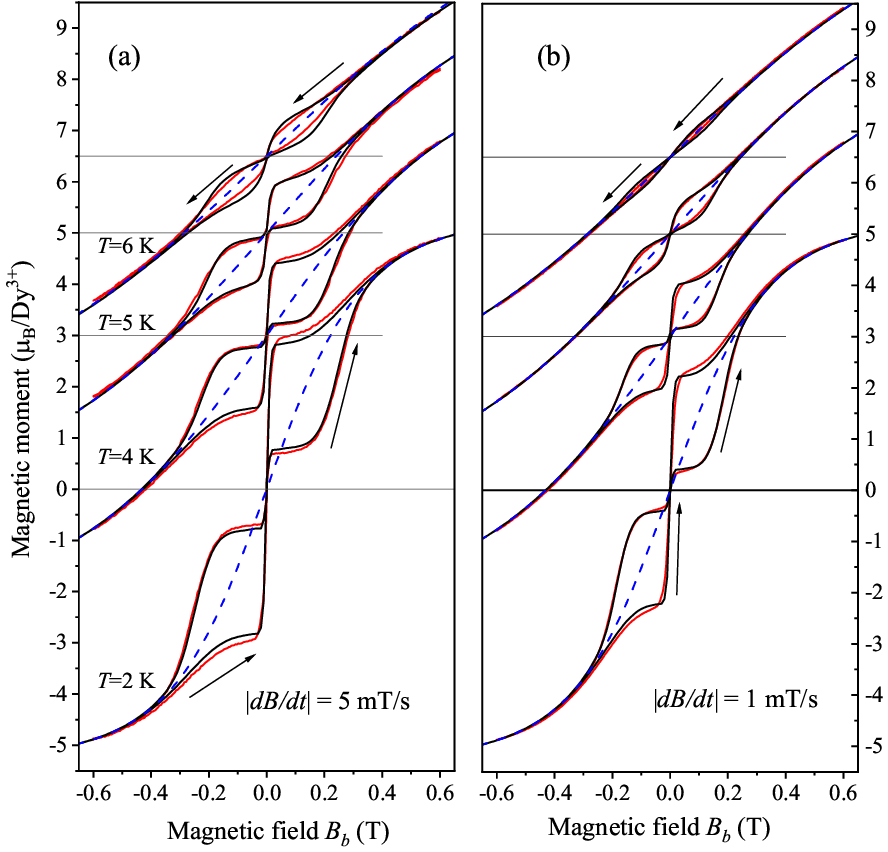}
\caption{\label{Fig1} Dynamic magnetization of the SrY$_2$O$_4$:Dy (0.01~at.\%) single-crystal in magnetic fields $\mathbf{B} \parallel \mathbf{b}$ for the sweeping rate of (a) 5 and (b) 1~mT/s at 2, 4, 5 and 6~K. Red and black lines represent the results of measurements and modeling. respectively. Dashed lines show the equilibrium magnetization. Magnetization curves for $T = 4,$ 5 and 6~K are shifted upward by 3, 5 and 6.5~$\mu_B$/Dy$^{3+}$, respectively; $\mu_B$ is the Bohr magneton.}
\end{figure*}

Figure~\ref{Fig1} shows the dynamic magnetization $M(B_b)$ measured on the single crystal SrY$_2$O$_4$:Dy$^{3+}$ (0.01~at.\%) in a swept magnetic field applied along the $b$-axis as well as the equilibrium one at temperatures 2, 4, 5 and 6~K ($B_{max} = 0.6$~T, $v = 5$ and 1~mT/s). 

In the magnetic fields oriented along the crystallographic axes, four Y1 sites, as well as four Y2 sites, in the unit cell are magnetically equivalent. According to the results of EPR measurements \cite{supp}),  the corresponding $g$-factors of the Dy$^{3+}$ ions are $g_{bb}(\mathrm{Dy1}) = 2.7$ and $g_{bb}(\mathrm{Dy2}) = 19.28$, so, the Dy2 ions provide the dominant contribution to the measured magnetization. 

\begin{figure}
\includegraphics[width=8.5cm, angle=0, clip=true]{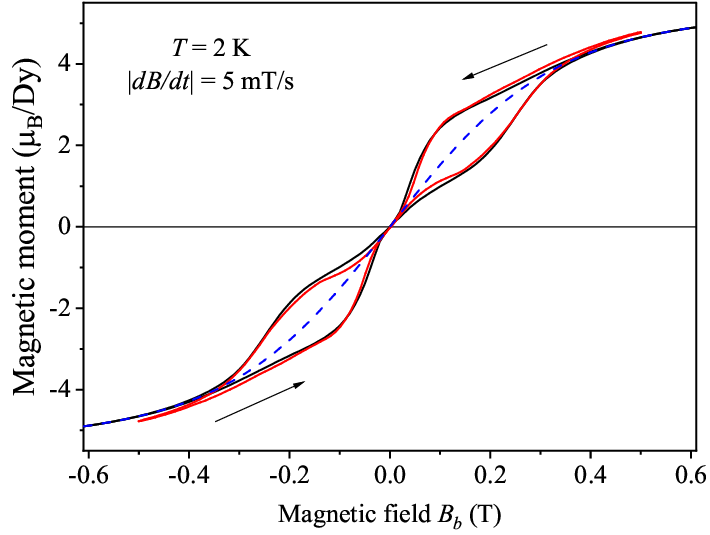}
\caption{\label{Fig2} Measured (red lines) and simulated (black lines) dynamic magnetization of the single-crystal sample SrY$_2$O$_4$:Dy (0.5~at.\%). Dashed line represents the equilibrium magnetization.}
\end{figure}

All registered dependencies $M(B_b)$ in the swept fields (Fig.~\ref{Fig1}) demonstrate well pronounced double-loop hysteresis. Loop areas reduce monotonously with decreasing a sweeping rate and increasing a temperature. We note a change of the loops shape from  quasi-rectangular upturns and downturns close to zero values of the applied field $B_b$ to a rounded spindle-type one between 5 and 6~K for decreasing sweeping rates in the range $5 - 1$~mT/s. A similar profile of the hysteresis loops was found in the dynamic magnetization of the SrY$_2$O$_4$:Dy single-crystal sample with much higher  dysprosium concentration (0.5~at.\%, Fig.~\ref{Fig2}).

Relaxation rates of the magnetization in SrY$_{2}$O$_{4}$:Dy$^{3+}$ single-crystals with different concentrations of impurity ions (0.01 and 0.5~at.\%) at a given temperature $T$ and fixed magnetic field $B_0$ along the $b$-axis were measured using fast switching of the magnetic field from the initial value $B_{in}$ in the equilibrium state of the sample to the $B_0$ value ($B_0 < B_{in}$ or $B_0  > B_{in}$). The magnetization evolution was tracked by periodic measurements (typically every second) after the target field $B_0$ had been set. Each magnetization was in fact an average over the integration time of the PPMS lock-in detector.

The measured magnetization evolution during the thermalization processes at the temperatures 2 and 4~K gives evidence for the two magnetic subsystems with fast and slow relaxion rates and strongly different contributions to the total magnetization. We identify these two subsystems with the Dy$^{3+}$ ions at Y1 and Y2 sites. In particular, the measured time dependencies of the magnetization after setting the field $B_0$, shown in Fig.~3 for $B_{in} = 0$, is described by the equation
\begin{equation}\label{Eqn1}
M(t)=\left[  M_1(B_0)+M_2(B_0)(1-e^{-t/\tau}) \right] /2,
\end{equation}
where $\tau$ is the relaxation time, $M_1(B_0)$ and $M_2(B_0)$ the equilibrium magnetic moments along the $b$-axis of Dy1 and Dy2 ions, respectively, at temperature $T$ and magnetic field $B_0$ (for example, at $T = 2$~K and $B_0 = 0.07$~T, $M_1(B_0) = 0.35$~$\mu_B/\mathrm{Dy1}$, $M_2(B_0) = 1.82$~$\mu_B/\mathrm{Dy2}$, $\tau = 1444$~s).

\begin{figure}
	\includegraphics{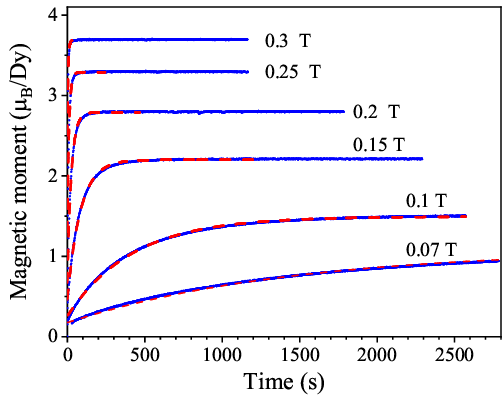}
	\caption{\label{Fig3} Registered (solid lines) relaxation to the equilibrium magnetization in SrY$_2$O$_4$:Dy (0.01~at.\%) sample after a fast setting of the external field $B_0$ along the $b$-axis at $T = 2$~K; for each curve, a value of $B_0$ is shown in the figure. Red dashed lines represent the fits of the data with single exponential slow evolution (\ref{Eqn1}) of the magnetization of Dy2 ions.}
\end{figure}

A small contribution of Dy1 ions to the total magnetization quickly achieves the equilibrium value $M_1(B_0)$, while variations of magnetic moments at Dy2 sites are well resolved and are successfully described by a single exponential relaxation model at relatively high magnetic fields $B_0 \geq 0.07$~T. Magnetic field dependence of the relaxation time $ \tau \propto B_0^{-4}$ at $T = 2 – 4$~K (see Fig.~4, the slope of the straight dashed line 1 equals $- 3.89$) evidences the dominant role of the direct spin-phonon relaxation process. However, when the target field $B_0$ tends to zero, the measured relaxation time of Dy$^{3+}$ ions shortens and reaches values as short as about 20~s ($T = 2$~K) at Y2 sites. This fact proves the existence of an additional relaxation mechanism effective in the region of weak magnetic fields. In the considered range of temperatures, Orbach–type relaxation is ineffective because of a large, 300~K, gap between the ground and the first excited CF sublevels of the ground multipet of the Dy$^{3+}$ ions at Y2 sites. However, the observed abrupt change of the shape of the hysteresis loops at the elevated temperatures of $5 - 6$~K as compared with the loops at $2 - 4$~K (see Fig.~\ref{Fig1}) evidences for the two-phonon Raman relaxation processes with the strong temperature dependence of the relaxation rate.   

\begin{figure}
	\includegraphics{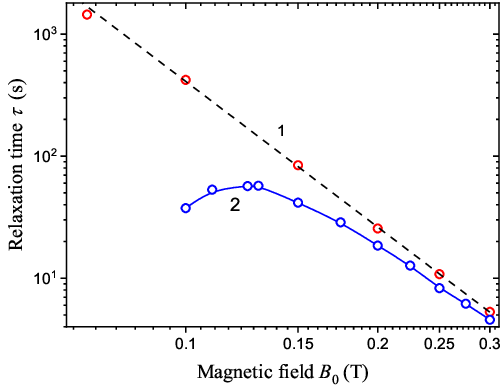}
	\caption{\label{Fig4} The relaxation time dependencies on the applied magnetic field at $T = 2$~K in SrY$_2$O$_4$:Dy crystals with Dy concentrations of 0.01~at.\% (1) and 0.5~at.\% (2). Line (2) is a guide for an eye.}
\end{figure}

Quantum tunneling and cross-relaxation processes affect the dynamic magnetization only in the relatively narrow range of sweeping magnetic field $- 0.02\ \mathrm{T} < B_b(t) < + 0.02\ \mathrm{T}$ comparable to widths of hyperfine structures of the ground Kramers doublets of odd dysprosium isotopes at Y2 sites. 

\section{\label{Sec3}MODELING OF THE OBSERVED HYSTERESIS LOOPS}
\subsection{Master equations}
Assuming a homogeneous distribution of impurity ions among Y1 and Y2 sites and neglecting interactions between the Dy$^{3+}$ ions in the strongly dilute samples, we describe the observed dynamic magnetization along the applied field by the quantum statistical expression
\begin{equation}\label{Eqn2}
\left\langle M_b(t) \right\rangle = \sum_{\lambda = Dy1,Dy2} \mathrm{Tr} \left[ M_{b,\lambda} \rho(\lambda,t)\right] / 2.
\end{equation}
Here $M_{\alpha.\lambda}$ are components of the magnetic moment operators of Dy$^{3+}$ ions, and $\rho (\lambda,t)$ is the single-ion density matrix of an impurity ion at the site Y1 ($\lambda = \mathrm{Dy1}$) or Y2 ($\lambda = \mathrm{Dy2}$) satisfying the generalized master equation \cite{rousochatzakis2005master} with the time dependent relaxation terms. To model the dynamic magnetization, we use the secular approximation that is adapted for the considered system; the non-diagonal elements of $\rho (\lambda,t)$ in the basis of eigenfunctions of the single-ion Hamiltonian $H_{\lambda}(t)$ are neglected. The diagonal elements $\rho_{nn} (\lambda,t) = \rho_n (\lambda,t)$ which determine populations of single-ion electronic (or electron-nuclear) states with energies $E_n(\lambda,t)$ satisfy the equations of motion
\begin{multline}\label{Eqn3}
\frac{\partial \rho_n}{\partial t} = \\ \sum_{k(k \neq n)}\left[ \rho_k W_{k \rightarrow n} (t) - \rho_n W_{n \rightarrow k} (t) + \Gamma_{kn}(t)(\rho_k - \rho_n) \right] + \\ 
\sum_{kpl(k \neq l, p \neq n)} W^{CR}_{n \leftarrow p,l \leftarrow k} (t) (\rho_p \rho_k - \rho_l \rho_n),
\end{multline}
here and below the site index $\lambda$ is dropped for simplicity. The time dependent coefficients at the right-hand side of Eq.~(\ref{Eqn3}) are transition probabilities between different states of a Dy$^{3+}$ ion induced by electron-phonon interaction ($W_{k \rightarrow n}$), LZS quantum tunneling ($\Gamma_{kn}$) and dipole-dipole interactions between paramagnetic ions (cross-relaxation (CR) processes with the probabilities $W^{CR}_{n \leftarrow p,l \leftarrow k}$ of simultaneous transitions in a coupled pair of ions, $p \rightarrow n$  in one ion and $k \rightarrow l$ in another ion \cite{bloembergen1959cross}).

Because of large gaps between the first excited and the ground state doublets of Dy$^{3+}$ ions at Y1 and Y2 sites, we can neglect populations of all excited CF energy levels at low temperatures 
($T \leq 6$~K). In the case of magnetically equivalent ions, the system of equations (\ref{Eqn3}) contains only two (even isotopes) or twelve (odd isotopes) equations for the relative populations of sublevels of the ground state doublet. The corresponding systems of differential nonlinear (in general case) equations with time dependent coefficients were solved numerically by the Newton method of successive approximations modified by varied steps of the sweeping field in searches for stable solutions. Calculations were performed separately for each of six contributions to the total dynamic magnetization from even and two odd isotopes, weighted in accordance with their natural abundances,  at sites Y1 and Y2. The initial step of increasing (decreasing) magnetic field had a value of $10^{-4}$~T. The normalization condition $\sum_n \rho_n = 1$ was checked at each step. The simulation starts from the numerical diagonalization of the single ion Hamiltonian $H(t=0)$ (see below) for an ion in the initial magnetic field $B_b(0) = B_{min}$ or $B_b(0) = B_{max}$ and the construction of the equilibrium density matrix $\rho_{nk}(0) = N \delta_{nk} \exp (-E_n(0) / k_B T)$ at a fixed temperature $T$ ($E_n(0)$ are eigenvalues of the Hamiltonian $H(0)$, $k_B$ is the Boltzmann constant and $N$ is the normalization factor).  

Transition probabilities in Eqs.~(\ref{Eqn3}) were calculated for fixed values of the magnetic field $B_b(t)$ using the single-ion Hamiltonian operating in the basis of electronic (for even Dy isotopes) or electron-nuclear (odd isotopes) states of the ground multiplet $^6 \mathrm{H}_{15/2}$. This effective Hamiltonian  
\begin{equation}\label{Eqn4}
H(t) = H_{\mathrm{CF}} + H_{\mathrm{Z}}(t) + H_{\mathrm{MHF}} + H_{\mathrm{QHF}}
\end{equation}
contains the CF interaction ($H_{\mathrm{CF}}$) defined by 15 CF parameters for each of two nonequivalent Y1 and Y2 sites determined from EPR and site-selective optical measurements \cite{supp}, magnetic dipole ($H_{\mathrm{MHF}}$) and electric quadrupole ($H_{\mathrm{QHF}}$) hyperfine interactions, and the Zeeman interaction ($H_{\mathrm{Z}}$).  The swept field is $B_b(t)$, and the steady transversal fields $B_a$ and $B_c$ are considered as the fitting parameters. We estimated the lower boundaries of the transversal fields $B_a = 0.284 \cdot 10^{-5}$~T and $B_c = 0.336 \cdot 10^{-5}$~T at Y2 sites from calculations of the mean-square dipolar fields \cite{prokof2000theory} of nuclear magnetic moments $\mathbf{m}_Y = g_n \mu_n \mathbf{I}$ of $^{89}\mathrm{Y}^{3+}$ ions  ($g_n = - 0.137$, $\mu_n$ is the nuclear magneton, $I = 1/2$),
\begin{equation}\label{Eqn5}
B_{\alpha} = \sum \dfrac{g_n \mu_n}{2r^5} \left[ \left( 3 \alpha^2 - r^2 \right) ^2 +
\left( 3 \alpha \beta \right) ^2 +\left( 3 \alpha \gamma \right) ^2 \right] ^{1/2} ,
\end{equation}
where summation was taken over all Y sites with coordinates $\alpha,\beta,\gamma$  ($\alpha \neq \beta \neq \gamma$) in the crystallographic frame with the origin at Y2 site. In the present work, the self-consistent description of the dynamic magnetization was achieved using the transversal fields $B_a = 1.45 \cdot 10^{-5}$~T and $B_c = 0.5 \cdot 10^{-5}$~T (note, these fields are comparable to the Earth magnetic field). In the crystallographic frame ($x \parallel a$, $y \parallel b$, $z \parallel c$),
\begin{equation}\label{Eqn6}
H_\mathrm{Z} = \mu_B g_J \left[ J_x B_a + J_z B_c + J_y B_b(t) \right] ,
\end{equation}
\begin{equation}\label{Eqn7}
H_{\mathrm{MHF}} = A_J \mathbf{J} \cdot \mathbf{I},
\end{equation}
\begin{multline}\label{Eqn8}
H_{\mathrm{QHF}} = B_Q \left\lbrace \left[ 3 J_z^2 - J (J + 1) \right] \left[ I_z^2 - I (I + 1) \right] / 3 + \right.\\
(J_x J_y + J_y J_x)(I_x I_y + I_y I_x) + (J_x J_z + J_z J_x)(I_x I_z + I_z I_x) + \\
\left. (J_y J_z + J_z J_y)(I_y I_z + I_z I_y) + (J_x^2 - J_y^2)(I_x^2 - I_y^2) \right\rbrace  + \\
P_0 \left[ 3 I_z^2 - I (I + 1) \right] + P_2 (I_x^2 - I_y^2) +P_{-2} (I_x I_y + I_y I_x) .
\end{multline}
Here, $J_{\alpha}$ and $I_{\alpha}$ are components of the total electronic angular and nuclear spin moment operators, respectively, $J = 15/2$ and $I = 5/2$. The effective Lande factor as well as the magnetic hyperfine coupling constants were slightly corrected as compared to the values corresponding to the Russell-Saunders approximation using the results of EPR measurements  \cite{supp}, $g_J(\mathrm{Dy1}) = 0.9925 \cdot 4/3$, $g_J(\mathrm{Dy2}) = 0.985 \cdot 4/3$, $A_J(^{161}\mathrm{Dy}) = - 3.683 \cdot 10^{-3}$~cm$^{-1}$,  $A_J(^{163}\mathrm{Dy}) = 5.163 \cdot 10^{-3}$~cm$^{-1}$). The upper three lines in Eq.~(\ref{Eqn8}) define contributions to the electric field gradient at the nucleus from the 4$f$ electrons localized on a rare-earth ion, the parameters $B_Q(^{161}\mathrm{Dy}) = 0.7021 \cdot 10^{-5}$~cm$^{-1}$ and $B_Q(^{163}\mathrm{Dy}) = 0.7319 \cdot 10^{-5}$~cm$^{-1}$ were calculated according to the corresponding definition \cite{abragam1970electron}. The lower line in Eq.~(\ref{Eqn8}) corresponds to the energy of the nuclear quadrupole moment at sites with the $C_s$ symmetry in the ionic lattice, the parameters $P_k$ were calculated using the nominal point ion charges, $P_0(^{161}\mathrm{Dy}1) = -1.092$, $P_2(^{161}\mathrm{Dy}1) = -2.543$, $P_{-2}(^{161}\mathrm{Dy}1) = -0.42$,  $P_0(^{161}\mathrm{Dy}2) = 1.935$, $P_2(^{161}\mathrm{Dy}2) = 11.615$,  $P_{-2}(^{161}\mathrm{Dy}2) = -0.164$ (in units of $10^{-5}$~cm$^{-1}$); for the $^{163}$Dy isotope, parameters $P_k(^{163}\mathrm{Dy}1)$ and $P_k(^{163}\mathrm{Dy}2)$ contain additional factor $Q(^{163}\mathrm{Dy})/Q(^{161}\mathrm{Dy}) = 1.043$. Note, the quadrupole hyperfine interaction affects weakly energies of hyperfine sublevels of Dy$^{3+}$ Kramers doublets, but it substantially increases the transition probabilities induced by the electron-phonon interaction (see Fig.~S7 in \cite{supp} and Ref.~\cite{taran2019role}).

\subsection{Electron-phonon relaxation}
At low temperatures, we can consider the interaction of 4$f$-electrons with long-wavelength acoustic phonons only. In this case, Hamiltonian of the electron-phonon interaction (EPI) operating in the space of electron-nuclear states of the ground multiplet $^6$H$_{15/2}$ is written as follows \cite{abragam1972spin,chudnovsky2005universal} (the second term at right hand side corresponds to the electron-rotational interaction)
\begin{equation}\label{Eqn9}
	H_{\mathrm{EPI}} = \sum_{\alpha,\beta} V_{\alpha\beta}e_{\alpha\beta} + i \left[ H,(\mathbf{J} + \mathbf{I}) \cdot \mathbf{\uptheta} \right]   ,
\end{equation}
where $H$ is the single-ion Hamiltonian (\ref{Eqn4}), $ e_{\alpha\beta} $ are components of the dynamic deformation tensor and $\mathbf{\uptheta}$ are vectors of dynamic rotations linear in phonon annihilation $a_{\mathbf{q}j}$ and creation $a^{+}_{\mathbf{q}j}$ operators:
\begin{equation}\label{Eqn10}
	e_{\alpha\beta} = \frac{1}{2} \sum_{\mathbf{q}j} Q_{\mathbf{q}j} \left[ \varepsilon_{j \alpha}(\mathbf{q}_0) q_{0 \beta} + \varepsilon_{j \beta}(\mathbf{q}_0) q_{0 \alpha} \right]   ,
\end{equation}

\begin{equation}\label{Eqn11}
	\theta_{\gamma} = \frac{1}{2} \sum_{\mathbf{q}j,\alpha\beta} Q_{\mathbf{q}j} e_{\alpha\beta\gamma} \varepsilon_{j \alpha}(\mathbf{q}_0) q_{0 \beta} .
\end{equation}
Here, $ Q_{\mathbf{q}j} = \left[ \hbar q / 2N m v_j (\mathbf{q}_0) \right] ^{1/2} (a_{\mathbf{q}j} + a^{+}_{-\mathbf{q}j}) $ , summations are taken over acoustic branches $ j $ of the vibrational spectrum and wave vectors $ \mathbf{q} $ of phonons with frequencies $ \omega_{\mathbf{q}j} = v_j (\mathbf{q}_0) q$; $ v_j (\mathbf{q}_0) $  is the sound velocity, $ \mathbf{q}_{0} $ and $ \mathbf{\upvarepsilon}_j (\mathbf{q}_0) $ are the unit propagation and polarization vectors, respectively, $ m $ is the cell mass, $ N $ is the number of cells, and $ e_{\alpha\beta\gamma} $ are components of the unit antisymmetric tensor. Parameters of the electron-deformation interaction $ b^k_{p,\alpha\beta} $ in the operators $ V_{\alpha\beta} = \sum b^k_{p,\alpha\beta} O^k_p $ ($ O^k_p $  are the Stevens operators) were computed within the exchange charge model (see Tables~S4 and S5 in \cite{supp}).

The single-phonon transition probability per unit time induced by the electron-phonon interaction (\ref{Eqn8}) is written as follows ($ E_k - E_n = \hbar \omega_{kn} > 0 $ )
\begin{multline}\label{Eqn12}
	W_{k \rightarrow n} = \frac{\omega_{kn}^3}{2 \pi \hbar d} \Bigg\langle \left| < k | \sum_{\alpha\beta} V_{\alpha\beta} \left[ e_{\alpha\beta} \right] + \right. \\ \left.
	i \hbar \omega_{kn} \sum_{\gamma} (J_{\gamma} + I_{\gamma}) [\theta_{\gamma}] | n > \right| ^2 \Bigg\rangle _{\mathrm{Av}} [n_0(\omega_{kn}) +1 ] ,
\end{multline}
where $ d $ is the crystal density, $n_0(\omega)$ is the phonon occupation number for $\omega_{\mathbf{q} j} = \omega $, $ [ e_{\alpha\beta} ] = [ \varepsilon_{j \alpha}(\mathbf{q}_0) q_{0\beta} + \varepsilon_{j \beta} (\mathbf{q}_0) q_{0\alpha} ] / 2$, $ [ \theta_{\gamma} ] = [ \varepsilon_{j \alpha}(\mathbf{q}_0) q_{0\beta} - \varepsilon_{j \beta} (\mathbf{q}_0) q_{0\alpha} ] / 2 $ , and symbol $ \left\langle \dots \right\rangle _{\mathrm{Av}} $ denotes the summation over acoustic phonon branches and averaging over directions of the propagation vector, for example, $ <[ e_{\alpha\beta}] [e_{\gamma\delta}] >_{\mathrm{Av}} = \sum_{j} \oint d \Omega \frac{[e_{\alpha\beta}] [e_{\gamma\delta}]}{4 \pi v_j(\mathbf{q}_0)^5}$ . The averaging procedures were carried out using the elastic constants $ C_{\alpha\beta\gamma\delta} $ of SrY$ _{2} $O$ _{4} $ \cite{gaillac2016elate} in equations which determine velocities and polarization vectors of sound waves ($ \sum_{\beta \gamma \delta} C_{\alpha\beta\gamma\delta} q_{0 \beta} q_{0 \gamma} \varepsilon_{\delta} = dv^2 \varepsilon_{\alpha} $). To illustrate the strong acoustic anisotropy of the crystal lattice, some results of calculations are presented below (in units of $ 10^{-18}$~(m/s)$ ^{-5} $):

$ \begin{array}{l l}
	<[e_{xx}][e_{xx}]>_{\mathrm{Av}}= 2.1213; &  <[e_{yy}][e_{yy}]>_{\mathrm{Av}} = 0.1516; \\  <[e_{zz}][e_{zz}]>_{\mathrm{Av}} = 0.3539; & \\
	<[e_{xy}][e_{xy}]>_{\mathrm{Av}} = 0.4203;  & <[e_{xz}][e_{xz}]>_{\mathrm{Av}} = 0.9221;  \\ <[e_{yz}][e_{yz}]>_{\mathrm{Av}} = 0.3900; & \\
	<[\theta_z][\theta_z]>_{\mathrm{Av}} = 0.4238; & <[\theta_y][\theta_y]>_{\mathrm{Av}} = 0.7176;  \\ <[\theta_x][\theta_x]>_{\mathrm{Av}} = 0.2950. &
\end{array} $

Computations of transition probabilities (\ref{Eqn12}), $ W_{k \rightarrow n} (t) $, for transitions between the sublevels of the ground doublet of Dy$ ^{3+} $ ions at Y1 and Y2 sites were carried out for each value of the field $ B_b(t) $ used in the simulations of the dynamic magnetization assuming the thermal equilibrium of a phonon bath (the corresponding occupation numbers equal $ n_0 (\omega) = [ \exp (\hbar \omega/k_B T) - 1]^{-1} $). It should be noted that the specific butterfly loops in molecular complex V$_{15}$ with the effective spin $S = 1/2$ ground state have been successfully described in Refs.~\cite{chiorescu2000butterfly,chiorescu2000non} taking into account the phonon bottleneck effect, however, a degree of resonant phonons heating depends strongly on the spin-phonon coupling strength and the concentration of paramagnetic ions.

The evolution of nonequilibrium populations of energy levels driven by the electron-phonon interaction is determined by the relaxation matrix $ W $, $ \partial \rho_n (t) / \partial t = \sum_k W_{nk} (t) \rho_k (t)$, where $ W_{nk} = W_{k \rightarrow n} $, $ W_{kn} = W_{nk} \exp (- \hbar \omega_{kn} / k_B T) $ and $ W_{nn} = - \sum_{k \neq n} W_{kn} $. The eigenvalues of the relaxation matrix determine the set of relaxation rates, in particular, for a two-level system, the relaxation time $ \tau = - (W_{11} + W_{22})^{-1} $. The calculated probabilities of the single-phonon transitions within the ground state doublet of the Dy$ ^{3+} $ ions at Y1 sites are large enough (in particular, for even isotopes in the field $B_b = 0.3$~T at $T = 2$~K, the relaxation time $\tau = 7.7 \cdot 10^{-4}$~s) to ensure quick thermalization of the Dy1 subsystem. Therefore, the simulated dynamic magnetization follows the Curie law in all the experiments performed in the present work. Thus, the Dy1 subsystem does not contribute to hysteresis loops due to the fast electron-phonon relaxation rates. For the Dy2 subsystem, we obtained about three orders of magnitude slower single-phonon relaxation rates (in particular, $\tau = 1.66$~s for even isotopes in the field $B_b = 0.3$~T at $T = 2$~K that is comparable to the measured relaxation time 5.3 s (see Fig.~\ref{Fig4}).

To fit the measured dynamic magnetization at elevated temperatures ($T = 5 - 6$~K), we supplemented the relaxation matrix of Dy$ ^{3+} $ ions at Dy2 sites with terms corresponding to the Raman relaxation processes which provide the relaxation rate of the ground Kramers doublet $\tau_R^{-1} = 7.2 \cdot (T/10)^9$~s$ ^{-1} $K$ ^{-9} $ at cryogenic temperatures \cite{chibotaru2018spin}. At  temperatures $T < 5$~K, these terms are practically ineffective, but at the temperature of 6~K the relaxation rate $ \tau_R^{-1} = 0.073 $~s$ ^{-1} $ exceeds remarkably the single-phonon relaxation rate 0.0174~s$ ^{-1} $ in the magnetic field $B_b = 0.1$~T and plays the dominant role in the region of lower fields providing the spindle-type shape of hysteresis loops. 

\begin{figure}
	\begin{minipage}[h]{0.87\linewidth}
		\includegraphics[width=1\linewidth]{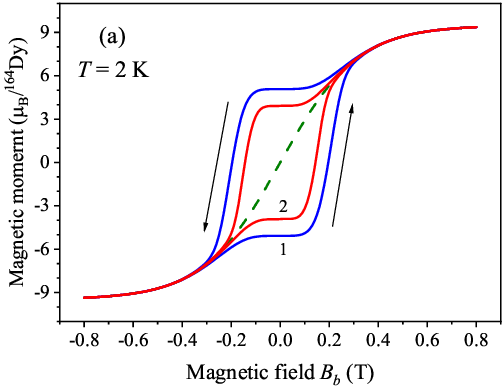}
	\end{minipage}
	\vfill
	\begin{minipage}[h]{0.89\linewidth}
		\includegraphics[width=1\linewidth]{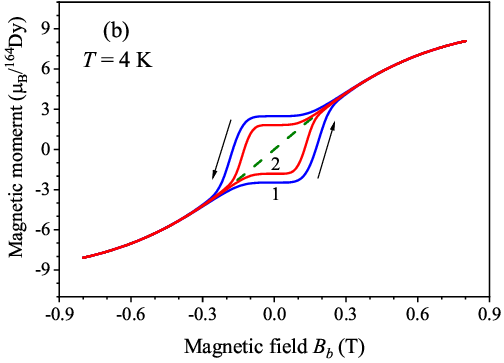}
	\end{minipage}
	\caption{Computed dynamic magnetization of even dysprosium isotopes at Y2 sites promoted by the electron-phonon interaction at temperatures (a) 2~K and (b) 4~K for sweeping rates of 5~mT/s (1) and 1~mT/s (2). Dashed lines correspond to the equilibrium magnetization.}
	\label{Fig5}
\end{figure}

Basing on numerical solutions of the master equations (\ref{Eqn3}) with an account for the electron-phonon interaction only, we have obtained the magnetic moments of Dy2 ions versus magnetic fields $B_b(t)$ dependencies that exhibit the single-loop hysteresis. For the temperatures of 2 and 4~K, the loops have similar shapes, and only the area of a loop diminishes with increasing the temperature and decreasing the sweeping rate (see Fig.~\ref{Fig5}). Clearly, the high-field magnetization behavior can be nicely described, however, the simulated low-field dependencies with the single hysteresis loop contradict experimental observations.

Transformation of a single-loop hysteresis to a double-loop one at cryogenic temperatures is induced by additional relaxation channels, namely, the cross-relaxation and the quantum tunneling, which are effective at low magnetic fields in the region of crossing hyperfine or Zeeman sublevels of the ground state Kramers doublet of Dy$ ^{3+} $ ions at Y2 sites.    

\subsection{Cross-relaxation}
\begin{figure}
	\begin{minipage}[h]{0.87\linewidth}
		\includegraphics[width=1\linewidth]{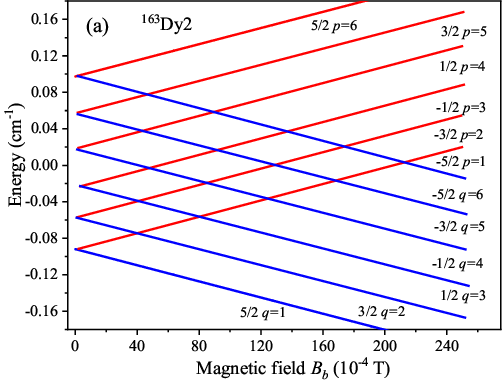}
	\end{minipage}
	\vfill
	\begin{minipage}[h]{0.89\linewidth}
		\includegraphics[width=1\linewidth]{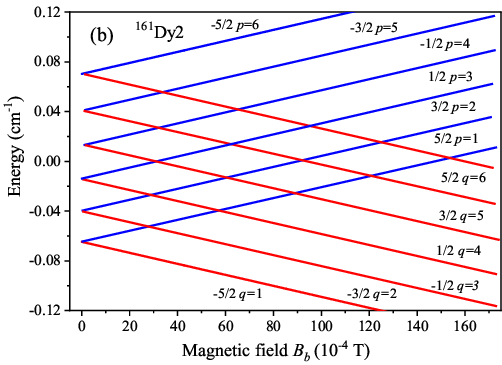}
	\end{minipage}
	\caption{Hyperfine structures of the ground doublets of (a)  $ ^{163}\mathrm{Dy}^{3+} $ and (b) $ ^{161}\mathrm{Dy}^{3+} $ ions at sites Y2 in weak magnetic fields $ B_b $ applied along the $ b $-axis. Indices $ p $ ($ q $) numerate hyperfine sublevels with increasing (decreasing) energies in the increasing field. }
	\label{Fig6}
\end{figure}

Energies of hyperfine sublevels of the ground doublet of Dy$ ^{3+} $ ions (odd isotopes) at Y2 sites versus the magnetic field $ B_{b} $ dependencies obtained by the numerical diagonalization of the corresponding single-ion Hamiltonian are presented in Fig.~\ref{Fig6}. Because of the strong Ising-type magnetic anisotropy, each hyperfine sublevel can be characterized by the nuclear spin projection ($I_y = -5/2, -3/2, \dots, 5/2$) on the applied field that has a maximum weight in the corresponding electron-nuclear wavefunction. Values of the magnetic fields $ B_b $ at the crossings of the hyperfine sublevels are presented in Table~\ref{Table1}. 

Cross-relaxation (CR) transitions lead to the formation of a nonequilibrium state of an ensemble of  magnetically equivalent multi-level paramagnetic ions (flip-flop transitions of magnetic moments affecting the magnetization involve at least three energy levels) or of two different magnetic subsystems with overlapping transitions in their responses to external time-dependent perturbations. Thus, the CR processes affect the dynamic magnetization of the two subsystems of odd isotopes of quasi-Ising Dy2 ions with approximately equal energy gaps between crossings of different pairs of hyperfine sublevels of the ground doublet at some fixed magnetic fields (see Fig.~\ref{Fig6}) and of the equilibrium and nonequilibrium subsystems of even isotopes at Y1 and Y2 sites, respectively, nearby the zero value of the sweeping field. The CR effects were considered qualitatively in Refs.~\cite{giraud2003quantum,wernsdorfer2002spin}.

The probability $ W^{CR}_{n \leftarrow p,l \leftarrow k} $ for the simultaneous transitions $k \rightarrow l$ of ion 1, and $ p \rightarrow n $ of ion 2 can be written as follows \cite{bertaina2006cross,malkin2008relaxation}:
\begin{equation}\label{Eqn13}
	W^{CR}_{n \leftarrow p,l \leftarrow k } = \frac{2 \pi}{\hbar^2} \left\langle | < n,l | H_{12} | p,k > |^2 \right\rangle _C \delta(\omega_{pn} - \omega_{lk}),
\end{equation}
where $ H_{12} $ is the Hamiltonian of interaction between the ions, $\hbar \omega_{pn}$ is the difference between energies of states $ p $ and $ n $, and $ \left\langle \dots \right\rangle _{C} $ denotes a configurational averaging over the distribution of the paramagnetic ions in the crystal lattice. 

\begin{table}
	\caption{\label{Table1} Magnetic fields $ B_b(p,q) $ (in units of $10^{-4}$~T) at the crossing points of the hyperfine sublevels ($ p $ and $q$) of the ground state doublet of $^{163}$Dy2 ions (I) and $^{161}$Dy2 ions (II) for zero transversal fields ($B_a = B_c = 0$).  See Fig.~6 for the values of the nuclear spin projections on the field $ B_b $ for electron-nuclear states $ p $ and $ q $.}
	\begin{tabular}{c | c c c c c c}
		$p/q$ \textbf{(I)} & 1 & 2 & 3 & 4 & 5 & 6 \\ \hline
		1 & 0 & 40.4 & 82.0 & 124.8 & 168.8 & 214.0 \\
		2 & -40.4 & 0 & 41.6 & 84.4 & 128.4 & 173.6 \\
		3 & -82.0 & -41.6 & 0 & 42.8 & 86.8 & 132.0 \\
		4 & -124.8 & -84.4 & -42.8 & 0 & 44.0 & 89.2 \\
		5 & -168.8 & -128.4 & -86.8 & -44.0 & 0 & 45.2 \\
		6 & -214.0 & -173.6 & -132.0 & -89.2 & -45.2 & 0 \\ \hline
		$p/q$ \textbf{(II)} & 1 & 2 & 3 & 4 & 5 & 6 \\ \hline
		1 & 0 & 28.2 & 57.6 & 88.1 & 119.8 & 152.6 \\
		2 & -28.2 & 0 & 29.3 & 59.9 & 91.6 & 124.4 \\
		3 & -57.6 & -29.3 & 0 & 30.5 & 62.2 & 95.0 \\
		4 & -88.1 & -59.9 & -30.5 & 0 & 31.6 & 64.5 \\
		5 & -119.8 & -91.6 & -62.2 & -31.6 & 0 & 32.8 \\
		6 & -152.6 & -124.4 & -95.0 & -64.5 & -32.8 &  0 \\ 
	\end{tabular}
\end{table}

Probabilities of CR processes depend strongly on the overlap of excitation energies of interacting ions. Note also that the cross-relaxation within the subsystem of magnetically equivalent two-level ions (even dysprosium isotopes) does not change the total magnetization. Thus, when considering the dynamic magnetization of the multisite SrY$ _{2} $O$ _{4} $:Dy$ ^{3+} $ crystals with strongly different $ g $-tensors of impurity ions at Y1 and Y2 sites, we can limit ourselves by taking into account the CR processes separately within the subsystems of $ ^{161} $Dy2 and $ ^{163} $Dy2 ions with the multi-level hyperfine structures of the electronic doublets, and the cross-relaxation between even isotopes of Dy1 and Dy2 ions in the magnetic field $ B_b(t) $ crossing the zero value. 
Taking into account finite widths of energy levels and assuming the homogeneous distribution of Dy$ ^{3+} $ ions over Y1 and Y2 sites, we can write the transition probability (\ref{Eqn13}) for ions belonging to the subsystems $ \lambda $ and $ \lambda^{\prime} $ in the following form \cite{wernsdorfer2002spin,bertaina2006cross} (we consider here the magnetic dipolar interactions between ions with magnetic moments $ \mathbf{M}_{\lambda} = g_{\lambda,J} \mu_B \mathbf{J}_{\lambda} $):
\begin{multline}\label{Eqn14}
	W^{CR,\lambda \lambda^{\prime}}_{n \leftarrow p,l \leftarrow k } = 2 \pi C_{\lambda^{\prime}} \frac{(\mu_B^2 g_{\lambda,J} g_{\lambda^{\prime},J})^2}{\hbar^2} \times \\
	\sum_{\alpha \beta \gamma \delta} g^{CR}_{\alpha \beta \gamma \delta} (\omega_{pn} - \omega_{lk}) k^{\lambda \lambda^{\prime}}_{\alpha \beta \gamma \delta} J^{(\lambda)}_{\alpha,np} J^{(\lambda^{\prime})}_{\beta,lk} J^{(\lambda)}_{\gamma,pn} J^{(\lambda^{\prime})}_{\delta,kl},
\end{multline}
where either $\lambda = \lambda^{\prime} = ^{161}$Dy1, $ ^{161} $Dy2, $ ^{163} $Dy1, $ ^{163} $Dy2, or $\lambda = ^{even}$Dy2 and $\lambda^{\prime} = ^{even}$Dy1, $C_{\lambda^{\prime}}$ is the concentration of $\lambda^{\prime}$-ions per Y lattice site, $g^{CR}_{\alpha \beta \gamma \delta} (\omega)$ is the CR form-function, $k^{\lambda \lambda^{\prime}}_{\alpha \beta \gamma \delta}$ are the lattice sums $k^{\lambda \lambda^{\prime}}_{\alpha \beta, \gamma \delta} = \sum_s a_{\alpha \beta} (\mathbf{R}_{\lambda \lambda^{\prime},s}) a_{\delta \gamma} (\mathbf{R}_{\lambda \lambda^{\prime},s})$ over sites of a Bravais lattice, and
\begin{equation}\label{Eqn15}
	a_{\alpha \beta} (\mathbf{R}_{\lambda \lambda^{\prime},s}) = \frac{1}{R^3_{\lambda \lambda^{\prime},s}} \left( \delta_{\alpha \beta} - 3 \frac{x_{\lambda \lambda^{\prime},s \alpha} x_{\lambda \lambda^{\prime},s \beta}}{R^2_{\lambda \lambda^{\prime},s}}  \right)  .
\end{equation}
Here, $ R_{\lambda \lambda^{\prime},s} $ is the radius-vector of the site s in the subsystem $ \lambda^{\prime} $ in the coordinate frame with its origin at the site belonging to the subsystem $ \lambda $. The computed lattice sums used in the modeling of the dynamic magnetization are presented in the Supplemental Material \cite{supp}.

The CR rates were calculated assuming the Gaussian line shape $g^{CR}_{\alpha \beta \gamma \delta} = \frac{1}{\sqrt{2\pi} \Delta} \exp [-(\omega_{pn} - \omega_{lk})^2 / 2 \Delta^2]$ of the spectral density of the energy reservoir corresponding to interactions between the Dy$ ^{3+} $ ions. The standard deviation $ \Delta $ of CR frequencies increasing with temperature and concentration of paramagnetic ions was estimated from the measured EPR linewidths  ($\Delta = 140 - 300$~MHz). The most important result of the CR processes is the appearance of the effective relaxation channels with the rates of up to $10^7$~s$^{-1}$ nearby the crossing points of the hyperfine sublevels of Kramers doublets in odd dysprosium isotopes. It should be noted that the calculated field dependencies of the CR-promoted relaxation rates might change remarkably if another CR line shape is used (the Lorentz distribution, in particular). 

\subsection{Quantum tunneling}
Computed values of gaps $G_{pq}$ (with a precision of $10^{-7}$~cm$ ^{-1} $) at the anticrossings of the hyperfine sublevels in the fields $B_b(p,q)$ are presented in Table~2. Transversal fields $B_a$ and $B_c$ of an order of $10^{-5}$~T practically do not change tunneling splittings $ G_{pq} $ in the spectra of odd isotopes but slightly shift $B_b$ values at all crossing points. Large gaps between the anticrossing hyperfine sublevels $ p $ and $ q $ of an order of $ 10^{-4} - 10^{-3} $~cm$ ^{-1} $ with the differences $ |m_p-m_q| = 1 $ between the corresponding nuclear spin projections on the swept field are induced by the magnetic hyperfine interaction (second order effects of an order of $ A_J^2/\Delta E $ where $ A_J $ is the magnetic hyperfine coupling constant and $ \Delta E $ is the CF energy of the first excited sublevel of the ground multiplet). Note, the direction of the applied field does not coincide with the quantization c-axis of the electronic angular momentum and is tilted by $\pm 9$~degrees in the $ (ab) $-plane from the principal directions of the $ g $-factor $ g_2 = 19.28 $ of Dy2 magnetically nonequivalent ions. 

An order of magnitude smaller gaps between the hyperfine sublevels with $ m_p = m_q $ are also induced by the magnetic hyperfine interaction. Additional significantly smaller (by about two orders of magnitude) gaps are induced by the quadrupole hyperfine interaction, but, as follows from calculations, this interaction additionally mixes electron-nuclear wave functions with nuclear spin projections $ m $ and $ m \pm 2 $ . This mixing provides weak additional anticrossings. The calculated gap between the sublevels of the ground doublet of even isotopes at Y2 sites arising from the transversal magnetic fields introduced above equals $ 3.6 \cdot 10^{-7} $~cm$ ^{-1} $, an order of magnitude less than the one communicated in Ref.~\cite{ding2018field}, in particular.

\begin{table}
	\caption{\label{Table2} Computed gaps $ G_{pq} $ (in units of $ 10^{-5} $~cm$ ^{-1} $) at anticrossing points in the hyperfine structure of the ground state doublet of $ ^{163} $Dy2 (I) and $ ^{161} $Dy2 (II) ions.}
	\begin{tabular}{c | c c c c c c}
		$p/q$ \textbf{(I)} & 1 & 2 & 3 & 4 & 5 & 6 \\ \hline
		1 & 0.00 & 0.00 & 0.04 & 0.16 & 14.83 & 7.46 \\
		2 & 0.00 & 0.05 & 0.07 & 18.75 & 4.00 & 54.34 \\
		3 & 0.04 & 0.07 & 19.88 & 1.22 & 68.65 & 0.60 \\
		4 & 0.16 & 18.75 & 1.29 & 72.78 & 0.27 & 0.13 \\
		5 & 14.83 & 4.00 & 68.65 & 0.27 & 0.16 & 0.00 \\
		6 & 7.46 & 54.34 & 0.60 & 0.13 & 0.00 & 0.00 \\ \hline
		$p/q$ \textbf{(II)} & 1 & 2 & 3 & 4 & 5 & 6 \\ \hline
		1 & 0.00 & 0.00 & 0.12 & 0.61 & 38.65 & 5.99 \\
		2 & 0.00 & 0.16 & 0.27 & 48.97 & 4.08 & 10.36 \\
		3 & 0.12 & 0.27 & 51.98 & 1.44 & 13.12 & 0.14 \\
		4 & 0.61 & 48.97 & 1.44 & 13.92 & 0.06 & 0.03 \\
		5 & 38.65 & 4.08 & 13.12 & 0.06 & 0.04 & 0.00 \\
		6 & 5.99 & 10.36 & 0.14 & 0.03 & 0.00 & 0.00 \\ 
	\end{tabular}
\end{table}

The analytical expression for the transition probability (per unit time) between anticrossing energy levels $ n $ and $ k $ with the minimal energy gap $\Delta_{kn}$ at the sweeping field $ B_b(t) = B_G $ 
\begin{equation}\label{Eqn16}
	\Gamma_{kn} (B_b) = \frac{\Delta_{kn}^2 \tau_{kn}}{2 \left\lbrace \hbar^2 + [g_{bb} \mu_{B} (B_b(t)-B_G)\tau_{kn}]^2 \right\rbrace }
\end{equation}
was derived accounting for the electron-phonon relaxation time $ \tau_{kn} = - (W_{kk} + W_{nn})/2 $ in Refs.~\cite{garanin1997thermally,villain1997effet,leuenberger2000spin}. The function (\ref{Eqn16}) has a Lorentzian shape with a width at half maximum   
\begin{equation}\label{Eqn17}
	\Delta B_{b,kn} = \frac{2 \hbar}{g_{bb} \mu_{B} \tau_{kn}}
\end{equation}
that does not depend on the gap $ \Delta_{kn} $. The measured relaxation time of Dy$ ^{3+} $ ions at Y2 sites at weak magnetic fields $|B_b(t)| < 0.02$~T  is not less than 20~s. In accordance with (\ref{Eqn17}), the width of the LZS regime does not exceed $ 6 \cdot 10^{-14} $~T and is much less than the minimal step (about $ 10^{-9} $~T) in the applied field variation used in simulations of the dynamic magnetization. As the duration  of the passage by the sweeping field through the quantum tunneling regions is much smaller than the relaxation time, we used the asymptotic LZS expression for nonadiabatic transition probabilities 
\begin{equation}\label{Eqn18}
	P_{nk} = \exp \left(- \frac{\pi \Delta_{nk}^2} {2 \hbar \mu_B g_{bb} |dB_b/dt|} \right)
\end{equation}
just after a passage of the field $ B_b $ through the anticrossing point. However, the corresponding evolution of the magnetization is strongly renormalized (see Fig.~S6 in \cite{supp}) by the CR processes effective within the magnetic field region of about $ 5 \cdot 10^{-4} $~T determined by the width of the CR form-function. Note, despite huge differences between the maximal values ($ 10^{11} – 10^{15} $~s$ ^{-1} $) of the quantum tunneling rate (\ref{Eqn16}), $ \Delta_{kn}^2 \tau_{kn} / 2 \hbar^2 $ , and the probabilities of the flip-flop CR processes, the accumulated corresponding changes of populations of the ground doublet sublevels have comparable values due to the inverse relation between the widths of the LZS and CR regimes.

Eventually, after taking into account the quantum tunneling and cross-relaxation, the simulated profiles of the hysteresis loops remarkably well reproduce the results of measurements at different temperatures and rates of the swept magnetic fields in SrY$ _{2} $O$ _{4} $ single-crystal samples doped with Dy$ ^{3+} $ (0.01 and 0.5~at.\%) ions (Figs.~\ref{Fig1} and \ref{Fig2}). As an example, separate contributions into the total magnetization from even and odd ($ ^{163} $Dy) isotopes at Y2 sites at the temperature 2~K for the sweeping rate 5~mT/s are shown in Supplemental Material \cite{supp} (Fig.~S8). The dynamic magnetization curves at low temperatures (2 and 4~K) in the increasing field from $ B_{min} = -B_{max} $ to $ B_{max} $ (as well as in the decreasing field from $ B_{max} $ to $ B_{min} $) show two quasi-plateaus at the entrance to and exit from the critical region of fast evolution of average magnetic moments per Dy$ ^{3+} $ ion determined by the widths of hyperfine structures of odd isotopes. When approaching this region, the value of the magnetic moment is determined by the electron-phonon relaxation rate. However, after passing through this region, the values of the magnetic moments are determined by the LZS tunneling renormalized by the CR processes. It should be underlined that the agreement of the performed modeling with the experimental data was achieved due to introduction of two fitting parameters, namely, components of the weak transversal magnetic field, comparable to the Earth magnetic field, affecting the impurity Dy$ ^{3+} $ ions at Y2 sites.

The unveiled low relaxation rates of impurity Dy$ ^{3+} $ ions at Y2 sites in single-ion magnets SrY$ _{2} $O$ _{4} $:Dy correlate with the unusual irreversibility of magnetic processes studied by means of the ultrasound technique in the concentrated SrDy$ _{2} $O$ _{4} $  single crystal at very low temperatures \cite{petrenko2017evolution}.

Recently, the hysteretic dynamic magnetization was observed at low temperatures in the paramagnetic phase of the concentrated inorganic dysprosium compounds DyScO$ _{3} $ ($ T_N = 3.1 $~K) \cite{andriushin2022slow} and LiDyF$ _{4} $ ($ T_N = 0.56 $~K) \cite{andreev2022first} where, most probably, formation of the hysteresis loops was caused by the magnetocaloric effect.  

\section{CONCLUSIONS}
The first observation and investigation of quantum tunneling of magnetization in the inorganic dilute rare-earth paramagnet, single-crystal of tetragonal double fluoride LiYF$ _{4} $ doped with non-Kramers Ho$ ^{3+} $ ions, was reported more than 20 years ago \cite{giraud2001nuclear,barbara2005quantum}. As follows from the results of our study, the multi-sublattice oxide SrY$ _{2} $O$ _{4} $ doped with Kramers Dy$ ^{3+} $ ions is the second  inorganic dilute rare-earth paramagnet that exhibits a hysteretic behavior of the dynamic magnetization similar to SMM.
  
The derived approach to the magnetization dynamics involved the comprehensive experimental studies of spectroscopic, magnetic and kinetic properties of the synthesized single-crystal samples with different concentrations of Dy$ ^{3+} $ ions, the subsequent analysis of the measured EPR and site-selective optical spectra and the magnetic relaxation rates in the framework of the semi-phenomenological crystal field model, and numerical solutions of the master equations.  This approach enabled us to successfully reproduce the hysteresis cycle shapes observed as well as their transformation with temperature, sweeping field rate and concentration of Dy$ ^{3+} $ ions. The important result of our work is the demonstration of the strong CR effects on the dynamic magnetization, in particular, the renormalization of the LZS incoherent transition probabilities at anticrossing points in the electron-nuclear manyfold of states in swept magnetic fields.

\begin{acknowledgments}
This work was supported by the Russian Science Foundation (project No.~19-12-00244). BZM is grateful to O.A.~Petrenko for useful discussions and to M.V.~Vanyunin for help in developing a Matlab code for solving a system of nonlinear equations of motion. The authors are grateful to M.A.~Cherosov for his assistance in magnetization measurements.
\end{acknowledgments}

\bibliographystyle{apsrev4-2}
\bibliography{SYO_Dy_Bib}

\end{document}